\documentclass[sigplan,10pt, screen]{acmart}

\AtBeginDocument{%
  }

\setcopyright{none}
\copyrightyear{}
\acmYear{}
\renewcommand\footnotetextcopyrightpermission[1]{}

\usepackage{amsmath}
\usepackage{booktabs}
\usepackage{multirow}
\usepackage{pifont}
\usepackage{tikz}
\usepackage{xspace}
\usepackage[normalem]{ulem} 
\usepackage{enumitem}

\usepackage{subcaption}

\newcommand{\name}{ShoreKeeper}

\newcommand{\sys}{\textsf{\name}\xspace}
\newcommand{\ours}{\textsf{\name}\xspace}
\newcommand{\cc}[1]{\texttt{#1}}

\renewcommand{\sectionautorefname}{§\kern-2.5pt}
\renewcommand{\subsectionautorefname}{§\kern-2.5pt}
\renewcommand{\subsubsectionautorefname}{§\kern-2.5pt}

\newcommand{\PP}[1]{\vspace{2px}\noindent{\bf #1.}}

\newcommand{\encircle}[1]{%
  \raisebox{.5pt}{\textcircled{\raisebox{-.9pt}{#1}}}}

\newenvironment{myitemize}%
  {\begin{itemize}
	[leftmargin=0cm,
		itemindent=.3cm,
		labelwidth=\itemindent,
		labelsep=0pt,
		parsep=0pt,
		topsep=2pt,
		itemsep=1pt,
		align=left]
  }%
  {\end{itemize}}
\newenvironment{myitemize2}%
  {\begin{itemize}
    [leftmargin=.3cm,
        itemindent=0pt,
        labelwidth=.15cm,
        labelsep=.15cm,
        parsep=0pt,
        topsep=2pt,
        itemsep=1pt,
        align=left]
  }%
  {\end{itemize}}

\newenvironment{myenumerate}%
  {\begin{enumerate}
	[leftmargin=0cm,itemindent=.5cm,labelwidth=\itemindent,
		labelsep=0pt,
		parsep=1pt,
		topsep=1pt,
		itemsep=3pt,
		align=left]
  }%
  {\end{enumerate}}

\fancypagestyle{submission}{%
  \fancyhf{}%
  \fancyfoot[C]{\thepage}%
}

\begin{document}
\raggedbottom
\title{\name: Accelerating Iterative Data Analytics in the Cloud via Distributed Recovery-Cost-Aware Caching}

\author{Jongseop Lee}
\affiliation{%
  \institution{POSTECH}
  \city{}
  \country{}
}

\author{Minsu Sun}
\affiliation{%
  \institution{POSTECH}
  \city{}
  \country{}
}

\author{Xinyue Ma}
\affiliation{%
  \institution{POSTECH}
  \city{}
  \country{}
}

\author{Atul Sandur}
\affiliation{%
  \institution{Amazon Web Services}
  \city{}
  \country{}
}

\author{Myeongjae Jeon}
\affiliation{%
  \institution{POSTECH}
  \city{}
  \country{}
}

\begin{abstract}
Large-scale data-intensive workloads powered by iterative graph or machine
learning processing have an insatiable demand for memory due to the
continuous generation of new intermediate data. Although existing caching
strategies strive to keep performance-critical data in memory, they often
fail in practice because they are limited to node-local memory management and
assume high node availability. In this paper, we introduce \ours, a framework
that maximizes execution efficiency by treating cluster-wide memory as a
unified, elastic pool. \ours employs a global optimization model that maps
distributed dataflows into a single graph abstraction to minimize the
recovery cost of data evicted from cache across the cluster. To handle memory
working sets and cluster capacities that fluctuate over time, \ours employs a
graph-pruning technique that permits the fast refinement of the caching plan.
These optimizations allow \ours to work seamlessly whether deployed on highly
available VMs or cost-effective preemptible instances in the cloud. Our
experiments show that \ours consistently outperforms state-of-the-art systems.

\end{abstract}

\maketitle
\pagestyle{submission}
\thispagestyle{submission}

\section{Introduction}

Numerous large-scale data-intensive workloads synthesize models to
characterize intricate real-world phenomena. For example, structural graph
analysis uncovers influential nodes based on the density and quality of their
incoming connections within massive networks~\cite{pagerank}.
Similarly, probabilistic forecasting or recommendation systems leverage
various machine learning algorithms to identify latent relational patterns
within vast datasets~\cite{sparklr, svdpp, als}. Such workloads rely on
iterative model refinement, with more iterations leading to a higher model
accuracy~\cite{graphx, mllib, systemml, iter3}. Modern distributed processing
frameworks typically express this refinement process using a dataflow-based
execution model~\cite{rdd, flinksite, googledataflow, nectar} that expands
incrementally as new computational iterations are initiated and generate
downstream \textit{data blocks}.

In-memory caching is vital for iterative data processing because it
eliminates recovery costs for intermediate data blocks accessed across
iterations. However, such computations constantly inflate the volume of
intermediate data as they progress through the dataflow, often straining
available memory resources~\cite{blaze, SparkSummit18_FB}. When faced with
memory shortage, an effective caching strategy must navigate a complex
trade-off between the execution gains of retaining individual data blocks and
the recovery costs---via recomputation or disk I/O---incurred if those blocks
are evicted. Earlier caching policies such as least reference count
(LRC)~\cite{lrc} and most reference distance (MRD)~\cite{mrd} bypass this
complexity simply by exploiting block dependency information exposed in the
dataflow. A more recent system, Blaze~\cite{blaze}, accounts not only for
dynamic access patterns, but also for the size and expected recovery cost of
each block. With this information, each compute node \emph{independently}
prioritizes caching of intermediate data that is more costly to retrieve at
each iteration boundary.

While effective at the individual node level, prior methods fall short of
optimizing caching decisions in diverse distributed settings.
In this work, we design and implement \ours, a generic cost-aware caching
system that works effectively for various deployment scenarios at scale.
Specifically, we address two critical challenges:

$\bullet$ \emph{Recovery cost imbalance.} Despite the efforts of system
schedulers to load balance across the cluster~\cite{spark, sparkscheduling},
new blocks generated by concurrent tasks across nodes are often
non-uniform in size and computation time.
Such imbalance occurs even when nodes execute the
dataflow with an identical sequence of operations, due to factors like
varying data filtering ratios or skewed key compositions after group-by-key
operations~\cite{skew1, skew2}. This poses two issues for node-level memory
management. (1) Some nodes may saturate their memory capacity earlier than
others, triggering premature block evictions while free space remains
available on remote nodes. (2) Even when no such free space exists, local
eviction decisions can lead to the loss of blocks that are globally expensive
to recover.
Without a cluster-wide view,
nodes cannot coordinate caching decisions to prioritize high-cost blocks
across the entire fleet.

\begin{figure*}[t]
  \centering
  \includegraphics[width=\linewidth,trim=0bp 10bp 10bp 4bp,clip]{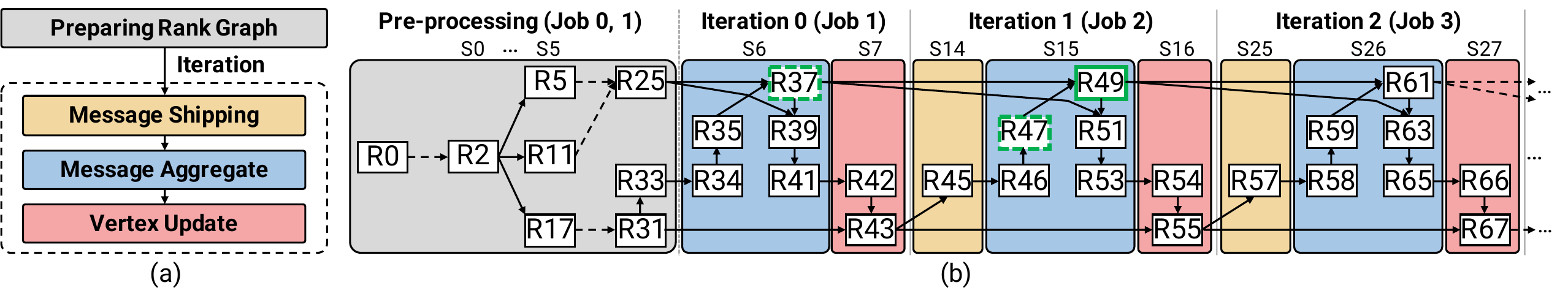}
  \caption{(a) A flowchart of the GraphX PageRank
    algorithm on Spark.
    (b) The corresponding DAG of the iterative
    execution.
    Each job (iteration) consists of stages
    (\textbf{Sx}: Stage \textit{x}), each producing intermediate states called RDDs
    (\textbf{Rx}: RDD \textit{x}). RDDs can be used across iterations (e.g., \textsf{R37} and \textsf{R43} in \textsf{Iteration 0}).
    Shuffle dependencies are omitted for clarity.}
  \label{fig:background-dag}
\end{figure*}

$\bullet$ \emph{Resource preemption.} To leverage scalability and ease of
maintenance, data analytics workloads often run in public clouds~\cite{emr,
dataproc, hdinsight}, using options ranging from on-demand virtual machines
(VMs) to preemptible instances~\cite{spot-instance, preemptive-vm, spot-vm}.
Node availability in such environments can fluctuate because of hardware or
software failures or, more commonly, due to preemptions that make room for
high-priority requests. Such disruptions risk the loss of crucial cached
blocks (costly to recover), prompting systems to maintain them across
constantly shifting memory pools. Cloud-native solutions do not currently
meet this goal: they simply attempt to use idle remote memory to temporarily
store random local blocks when preemption notices
occur~\cite{spark-decommission}.

\ours addresses these issues by introducing a global, cluster-level cost
optimization framework aimed at minimizing application execution time subject
to the \emph{time-varying} aggregate memory capacity of the cluster. At its
core, the system executes a mixed integer linear programming (MILP) solver
that determines not only \emph{which} data blocks to keep in memory
(caching), as prior cost-aware systems do, but also \emph{where} each data
block should reside across the cluster (placement). \ours treats every
executor as a candidate location and precisely weighs the gain of retaining a
partition against the cost of evicting resident data at the destination. To
explore such plans, \ours transforms the distributed dataflows across compute
nodes into a single unified computational graph that captures cluster-level
recovery costs and accounts for cross-node transfer latency.

\ours applies a dataflow
pruning algorithm to keep MILP solving tractable. It prunes the optimization
scope to only those blocks relevant to upcoming iterations in a near-future
window, reducing decision latency to under 4 secs in our experiments. This latency is low enough to invoke
the solver on-the-fly
for intermediate data blocks produced at every iteration at scale. For
blocks designated for another node, \ours performs cross-node transfers in
the background, using only the network capacity that foreground computation
leaves idle.

\ours's strategy for preemptible instances
exploits cloud-native preemption events to dynamically refine its caching
plan under reduced cluster capacity. Most cloud providers issue preemption
alarms with predefined \emph{grace periods} (e.g., 30 secs in Google Cloud).
Upon receiving an alarm, \ours re-triggers its solver to quickly migrate
critical blocks from nodes slated for reclamation within the grace period.
However, preemption alarms can occasionally be missed in
practice~\cite{awsspotnotice, bamboo}. To handle
such misses, \ours
periodically replicates a minimal set of blocks produced by time-consuming
shuffle communication.

We build \ours atop Apache Spark 3.3.2~\cite{spark} and evaluate it against
state-of-the-art in-memory caching~\cite{lrc, mrd, blaze} and storage-based checkpoint systems~\cite{flint} over
four benchmarks. With on-demand instances, \ours achieves up to 6.11$\times$
speedup over cost-agnostic caching methods  and up to
2.05$\times$ over Blaze~\cite{blaze}, with the largest gains on workloads
with the highest cross-executor imbalance. With preemptible instances, the
performance gap widens further: \ours achieves up to 2.87$\times$ speedup
over Blaze combined with Spark's Decommissioner~\cite{spark-decommission}, as
cost-aware evacuation and proactive replication protect high-cost partitions.

\section{Caching for Iterative Dataflow Execution}
\label{ss:dag-execution}

Distributed processing frameworks typically express the dataflow of
iterative, data-intensive computations as a directed acyclic graph
(DAG)~\cite{spark, flink, dryad}. \autoref{fig:background-dag} illustrates a
simplified PageRank~\cite{pagerank} DAG on Apache Spark~\cite{rdd, spark} as
an example. In this DAG, vertices represent intermediate data blocks termed
resilient distributed datasets (RDDs), and edges represent the
transformations that derive downstream RDDs from upstream ones. An
\emph{iteration} corresponds to a single data-processing job, the unit of
scheduling across cluster resources. Each job is divided into \emph{stages}
with boundaries defined by shuffle operators (e.g., group-by-key, join) that
redistribute data across the network by element key. Within a stage, each RDD
is partitioned and processed in parallel by tasks across distributed
processes termed \emph{executors}, each with a fixed memory allocation. As
shown in~\autoref{fig:background-dag}, this DAG structure repeats across
iterations, with each job consuming the output partitions of its
predecessors, making effective caching critical for performance.

When memory is full, cached partitions are either spilled to disk or
discarded. While a spilled partition is restored via disk I/O, recovering a
discarded partition requires recomputing it from the lineage, which is the
recorded sequence of transformations starting from the source datasets
\cite{rdd}. The cost of recomputation depends on the memory-residency state
of every partition along its lineage~\cite{blaze, lrc, mrd}. For instance,
\textsf{R49} in~\autoref{fig:background-dag} depends on multiple upstream
partitions: \textsf{R47} from the same iteration and \textsf{R37} from the
previous one. If both are in memory, recovering \textsf{R49} requires a
single transformation. If both have been discarded, the executor must
recursively recompute all missing partitions along the entire path back to
the original input. This cascading recomputation significantly amplifies the
total recovery cost for a single cache miss by orders of magnitude.

\PP{Cost-Aware Caching} Heuristic-based schemes, ranging from classic Least
Recently Used (LRU)~\cite{spark} to dependency-aware policies such as Least
Reference Count (LRC)~\cite{lrc} and Most Reference Distance
(MRD)~\cite{mrd}, establish caching priorities based on historical usage
patterns or partition dependencies in the DAG. However, none of these
policies account for the actual recovery costs associated with future
accesses to evicted partitions. Cost-aware caching methods address this gap
by explicitly modeling the recovery cost of each partition when making
eviction decisions. Blaze~\cite{blaze}, the state-of-the-art in this family,
estimates the recovery cost $cost(p_i)$ of partition $p_i$ as the minimum of
the disk access cost $cost_d(p_i)$ and the recomputation cost $cost_r(p_i)$:
\noindent
\begin{align}
  cost_d(p_i) &=
    size(p_i) ~/~ \textit{throughput}_\textit{disk}
  \label{eq:blaze-disk} \\
  cost_r(p_i) &= \textstyle\max_{p_k \in A_i}
    ((1 - m_k) \cdot cost(p_k) + cost_{k \to i})
  \label{eq:blaze-recomp}
\end{align}
\noindent
where $A_i$ is the set of parent partitions of $p_i$, $m_k$
indicates whether parent $p_k$ resides in memory, and $cost_{k \to i}$ is the
computation time for deriving $p_i$ from $p_k$.
With these estimates, Blaze formulates caching as an integer linear
programming (ILP) problem that minimizes the total potential recovery cost of
the set of partitions $J$ accessed by upcoming iterations, subject to a
memory capacity constraint:
\begin{alignat}{2}
  &\text{Minimize} \;\; && \textstyle\sum_{p_j \in J} (d_j \cdot cost_d(p_j) + u_j \cdot cost_r(p_j))
    \label{eq:blaze-objective}
\end{alignat}
\noindent
where $d_j$ indicates whether partition $p_j$ in $J$ has been
spilled to disk, and $u_j$ indicates whether it has been discarded. By
unifying these cost factors into a single objective, Blaze can systematically
balance caching benefits against recovery costs. This approach identifies
optimal plans that outperform cost-agnostic heuristics by up to
2.86$\times$~\cite{blaze}.

\section{Limitations and Insights}
\label{ss:executor-level-limits}

Despite the performance gains of cost-aware caching, existing solutions focus
primarily on node-local memory management. Specifically, they require each
executor to make independent caching decisions based solely on its local
partition set and memory budget. In this section, we characterize iterative
data processing in
public cloud settings and analyze two sources of dynamism that challenge
current designs.

\subsection{Recovery cost imbalance}
\label{ss:data-skew}

Iterative workloads often process data with non-uniform distributions. In
graph workloads, for instance, a small fraction of vertices may account for a
disproportionate share of edges. Offline partitioning methods address this
prior to execution: vertex-cut partitioning~\cite{powergraph} distributes edges evenly across
executors by replicating high-degree vertices,
and GraphX~\cite{graphx}
minimizes the number of such replicas. Together, these strategies yield nearly balanced per-executor partition sizes at the start of execution.

However, the initial balance does not always persist
as execution progresses.
The variation in memory arises from runtime factors (e.g., selective
filtering, aggregation fan-in, and shuffle redistribution)~\cite{skew1,
skew2, skew3} that offline partitioning methods cannot anticipate.
Because each operator consumes the output of a preceding stage or iteration,
uneven partition sizes propagate into uneven computation times, even though
every executor applies the same sequence of operations.

As iterations unfold, new intermediate states accumulate and add memory pressure. In distributed cloud settings, evicted partitions from executors under such high pressure have an opportunity to be migrated to a remote executor's spare memory, stored in local disk or discarded so they can be recomputed when needed. Persisting a partition or discarding it are favorable options if caching in a remote executor for later retrieval can result in high network transfer costs. On the other hand, processing partitions on the same executor can result in compute hotspots. Since the overall iteration time is bound by the slowest executor, such bottlenecks can significantly impact the job completion time.

\begin{figure}[tbp]
  \centering
  \includegraphics[width=\linewidth]{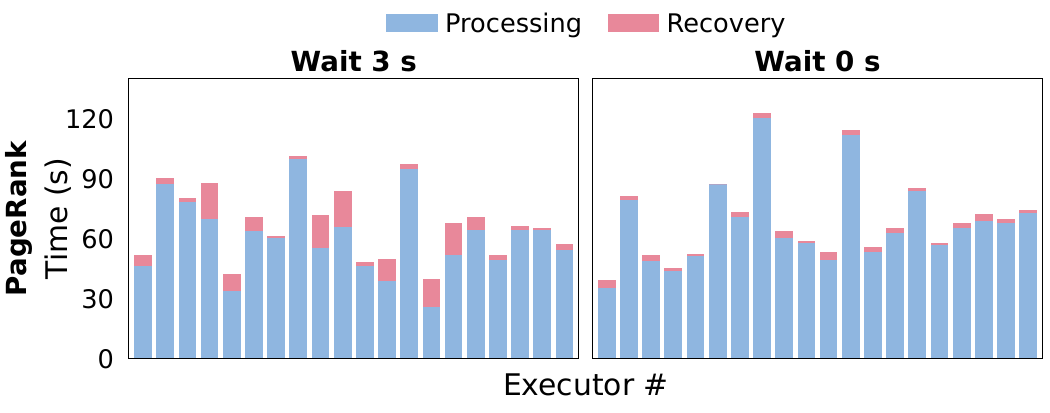}
  \caption{Per-executor processing
    and recovery costs under Blaze 
    with 3~s (left) and 0~s (right) locality waits.
    }
  \label{fig:pr-locality-wait}
\end{figure}

Our key insight is that end-to-end job completion time depends on these factors: (1) cost of task computation running on executors, (2) caching strategy which identifies important partitions incurring high recomputation cost (if discarded) or disk access cost (if persisted), and (3) remote fetching cost based on the distribution of dependent partitions across executors. These factors are inter-dependent and impact each other. Consider locality-aware data analytics schedulers~\cite{spark}, which reduce remote fetching cost but amplify memory imbalance by directing tasks toward executors that already hold the most data, thereby increasing compute pressure. On the other hand, load balancing schedulers~\cite{delayscheduling} can avoid compute hotspots but incur high network transfer costs to fetch remote partitions. \autoref{fig:pr-locality-wait} confirms the tradeoff between waiting too long (e.g., 3 seconds) for an executor to have compute cycles, resulting in high recovery costs vs. distributing the partitions (e.g., wait time of 0 seconds) resulting in potentially long remote fetch times (included in the processing time with blue bars). 

\begin{figure}[tbp]
  \centering
  \includegraphics[width=\linewidth,trim=4bp 10bp 5bp 11bp,clip]{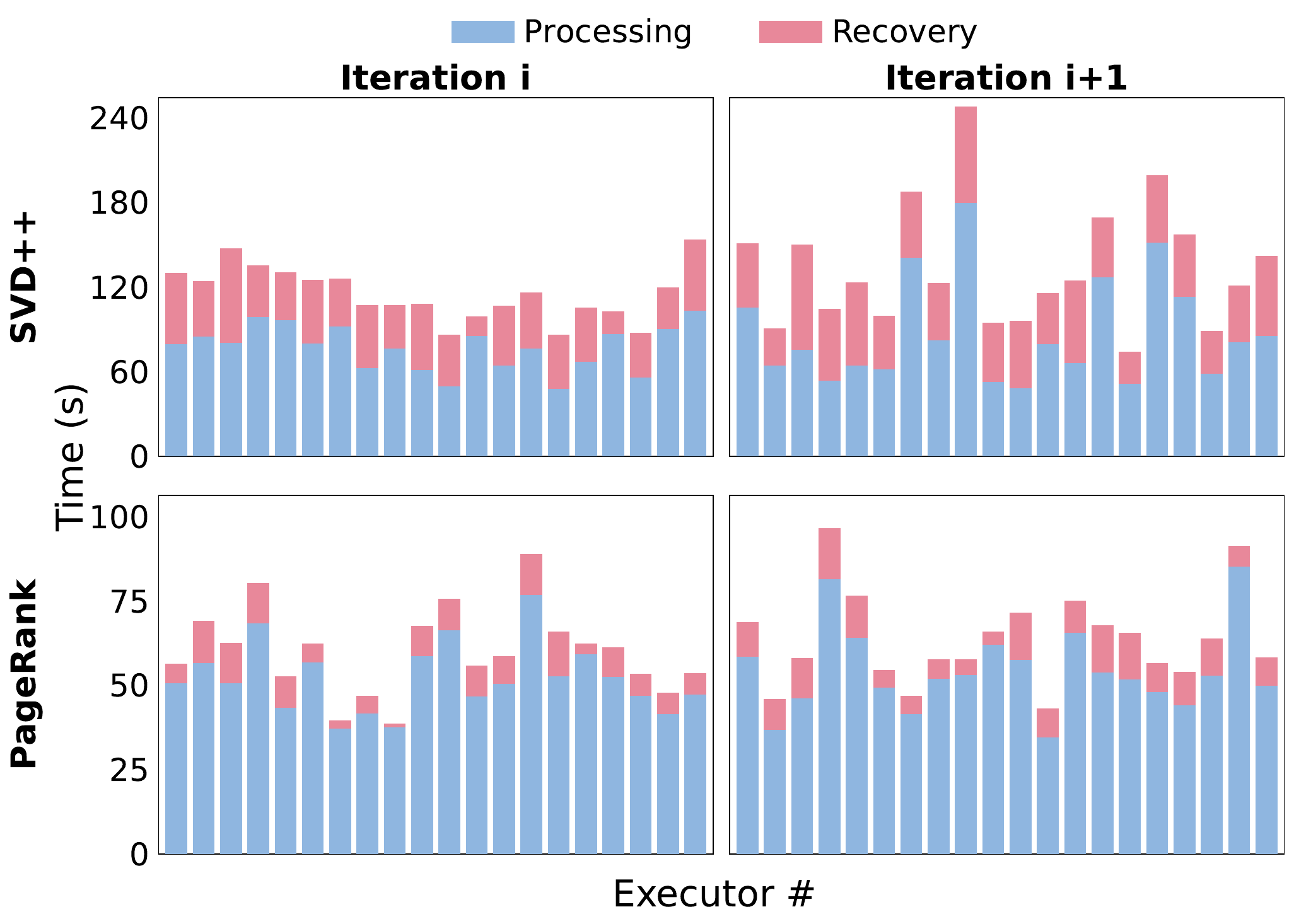}
  \caption{Per-executor processing (in blue) and recovery costs (in red) under Blaze for SVD++ (top) and GraphX PageRank (bottom) over two consecutive iterations.
    }
  \label{fig:comp-skew}
\end{figure}

Critically, as \autoref{fig:comp-skew} shows, the distribution of recovery
costs shifts across consecutive iterations as iterative computation reshapes the intermediate data, making
the set of heavily affected executors non-static.
This pattern holds for both PageRank and SVD++, indicating that recovery
imbalance is structural rather than workload-specific. Prior runtime
methods attempt to close this gap by redistributing data records in RDDs
regularly at iteration boundaries~\cite{skew1, skew2, skew3}. However, these
methods target computation and shuffle balance without accounting for
caching/recovery cost-aware data placement.
This creates an opportunity: executors with
lower iteration time and recovery cost can absorb
partitions from heavier peers to reduce iteration time.

\subsection{Deployment Environments}
\label{ss:transient-instance}

Large-scale data analytics
increasingly run on public clouds to leverage their scalability and ease of
maintenance~\cite{moirai, lakehouse, skypilot}. Such deployment choices in
the cloud add another layer of dynamism to
iterative workloads, beyond the recovery cost imbalance discussed above.
This work targets two widely used resource tiers: high-availability on-demand
virtual machines (VMs) and cost-effective preemptible
instances~\cite{spot-instance, preemptive-vm}. Given these options, users
typically compose compute fleets using either resource type or a mix of
both~\cite{cantbelate, harvest, emr_spot}. Since preemptible instances are
allocated on a best-effort basis, the instance pool undergoes both
preemptions and new allocations
driven by high-priority demands (e.g., VM requests). Therefore,
frameworks operating on preemptible instances experience highly dynamic
compute and memory capacity during execution.

\begin{figure}[tbp]
  \centering
      \includegraphics[width=\columnwidth,trim=5bp 7bp 4bp 4bp,clip]{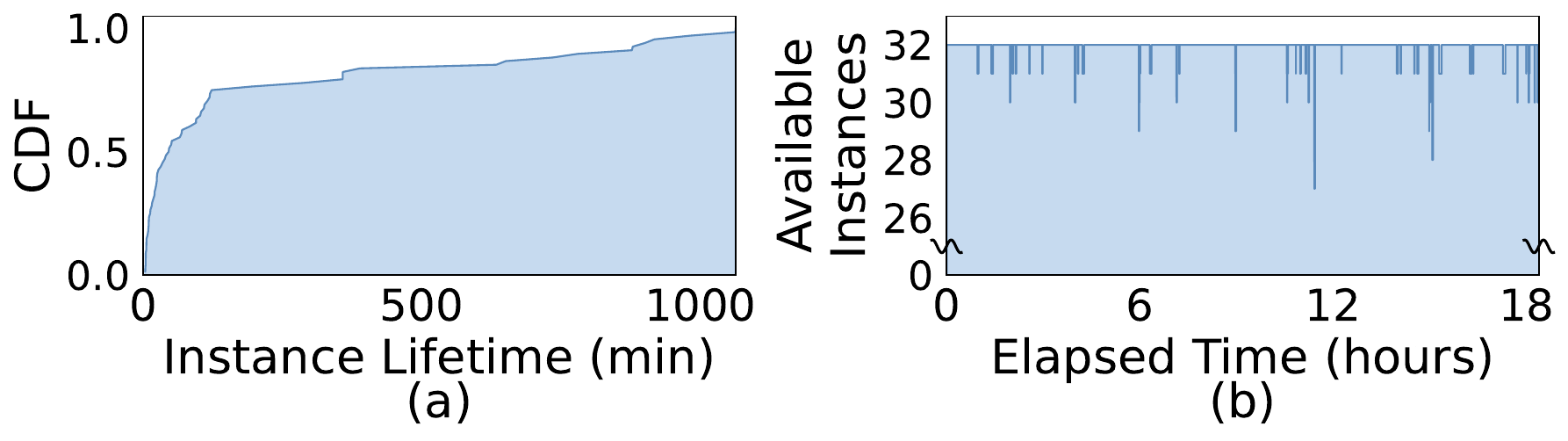}
  \caption{Preemption patterns from
    18-hour
    AWS trace with 32 \cc{r5.xlarge} Spot
    instances, \cc{us-east-1}.
    (a) CDF of instance's lifetime.
    (b) Number of active instances.
    }
  \label{fig:spot-trace}
\end{figure}

We characterize preemption patterns from an 18-hour trace collected on 32
\cc{r5.xlarge} AWS Spot Instances in \cc{us-east-1} from March 28 to 29,
2026. The CDF of instance lifetimes (\autoref{fig:spot-trace}a) shows that
approximately 25\% of instances are preempted within 30~minutes of launch.
\autoref{fig:spot-trace}b further confirms that preemption events occur
frequently throughout the trace, with the number of active instances
repeatedly dropping below the initial 32. Given this
preemption rate, iterative workloads that run for tens of minutes to hours
are
likely to encounter at least one preemption during execution.

\begin{figure}[tbp]
  \centering
      \includegraphics[width=\columnwidth,trim=4bp 7bp 5bp 9bp,clip]{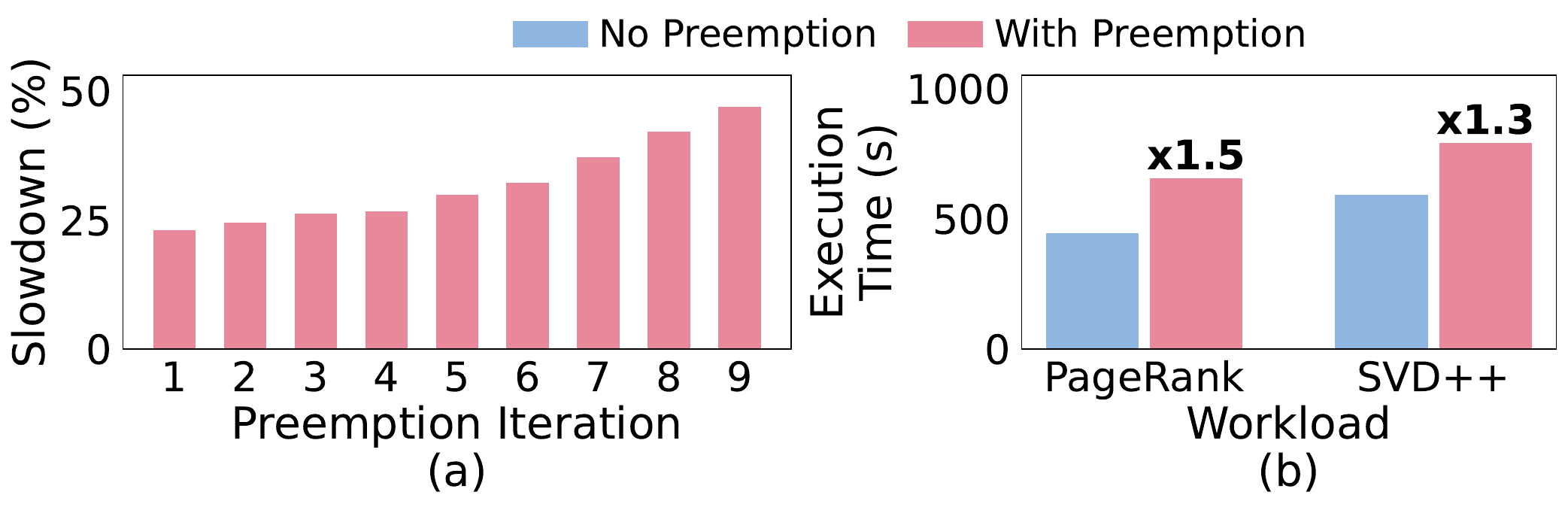}
  \caption{Impact of a single node preemption (10 iterations).
    (a) Slowdown of PageRank when one of 10 executors
    is preempted at different iteration points.
    (b) End-to-end completion time 
    with and without
    preemption at the last iteration.
    }
  \label{fig:preemption-impact}
\end{figure}

\PP{Impact of Node Departure and Arrival} When a compute instance departs,
all in-memory cached partitions accumulated over multiple iterations on that
node are lost,
forcing recovery through lineage recomputation or disk I/O under reduced
cluster capacity. The performance impact of this loss is highly sensitive to
when it occurs. As iterations proceed and more intermediate data accumulates,
the recovery lineage grows deeper, leading to a steady increase in completion
time for later preemptions (\autoref{fig:preemption-impact}a). Losing a
single instance during the final iteration increases completion time by
30--50\% for the workloads evaluated in \autoref{fig:preemption-impact}b.

When a new instance joins, it provides no immediate caching benefit by
default because cache population is lazily coupled with tasks subsequently
scheduled on that node. To our knowledge, no mechanism exists to
\emph{proactively} redistribute cached partitions across the enlarged
instance pool.

\PP{Cloud-Native Support} To mitigate preemption impacts, cloud providers
issue interruption alarms with a brief grace period (e.g., 2 mins on AWS, 30
secs on Google Cloud). Managed services such as AWS EMR~\cite{emr_spot} and
Google Cloud Dataproc~\cite{dataproc} leverage these alarms to trigger
graceful decommissioning, aiming to safely offload stateful data before a
node departs. The Spark Node Decommissioner~\cite{spark-decommission} exemplifies this
approach: upon receiving an alarm, the departing executor migrates cached
partitions to surviving peers. However, it selects partitions at random
(oblivious to recovery costs),
and the total volume is strictly constrained by the idle memory available in
peers. In practice, such headroom is rarely sufficient, as executor memory is
often heavily consumed by intermediate data accumulated across iterations.

Preemption alarms are not always guaranteed in practice~\cite{awsspotnotice,
bamboo}. Unalarmed departures bypass decommissioning entirely, causing the
loss of all cached data on the affected node. However, our traces show an
extremely low alarm miss rate (0.3\%), indicating that periodic checkpointing
to persistent storage (e.g., AWS EBS)~\cite{flint, phoebe} is impractical for
such rare events, given its substantial overhead~\cite{salama-ft, ftopt} (see
\autoref{fig:preemption-main}). This motivates a lightweight, proactive
approach
tailored to rare alarm misses.

\section{\ours Overview}
\label{sec:overview}

\ours is a cost-aware caching system that manages aggregate memory across
on-demand VMs \emph{or} preemptible instances as a
unified pool\footnote{\ours's current design focuses on clusters provisioned
with a single resource type. We discuss the case for combining resource types
in \autoref{sec:discussion}.}. \ours has the following design goals to
overcome the limitations identified in
\autoref{ss:executor-level-limits}.

\begin{myitemize2}

\item \textbf{Cost-aware cluster-wide placement.} To mitigate
    severe imbalance in recovery costs across executors,
    \ours must place each partition on the executor
    that minimizes cluster-wide recovery cost, rather than confining it to
    its origin.
    \ours schedules cross-executor transfers carefully to
    avoid interference with foreground network operations.

\item \textbf{Cost-aware preemption handling.} On receiving
    a preemption alarm, \ours must prioritize the migration of partitions with
    high recovery costs over random ones. These migrations are executed
    within rigid grace periods while minimizing performance damage on peer
    nodes even under significant memory pressure.

\item \textbf{Fast planning.} Caching and placement decisions must
    quickly respond to frequently changing workload memory
    states and cluster capacity. \ours aims to compute
    plans within
    seconds to keep up with typical iteration times and rigid grace periods
    imposed by cloud providers.

\item \textbf{Alarm miss handling.} \ours employs minimal
    in-memory replication to handle rare alarm misses. This strategy
    should avoid high performance overhead, given that alarms are rarely missed.

\end{myitemize2}

\subsection{Design Overview} \label{ss:design-overview}

While features specific to resource preemptions (e.g., cost-aware migration
and replication) are inactive on stable on-demand VMs, \ours provides a
unified solution rather than fragmented fixes. \ours further simplifies
its operations through a clean separation of planning and execution.

\PP{Unified Planner (\autoref{s:cache-planning})} \ours orchestrates caching,
placement, and replication through a unified optimization framework.
It maintains accurate cost estimates via a cost model informed by both
offline and online profiling, and solves a mixed integer linear programming
(MILP) to
derive the cluster-wide
plan. \ours jointly determines \emph{whether} to cache a partition,
\emph{where} to place it, and \emph{whether} to replicate it, rather than
treating these decisions in isolation.
The solver considers every executor as a candidate location for every
partition while incorporating the transfer cost into the placement decision.
When placing a partition on a non-origin executor, \ours weighs
the gain of retaining that partition against the cost of evicting resident
data at the destination.

\PP{Runtime Manager (\autoref{s:runtime-manager})} \ours invokes the planner
(1) at periodic iteration boundaries, (2) upon preemption alarms, and (3)
when a new executor joins the cluster. The Runtime Manager translates each
new plan into per-executor actions.
It diffs each new plan against the current placement, so only partitions
whose placement changes are transferred. These transfers run asynchronously,
batched per destination, and consume only the network capacity that
foreground execution (e.g., shuffle stages) leaves idle.
On receiving a preemption alarm, \sys migrates partitions in descending order
of recovery cost, transferring as much critical data as possible within the
grace period.

\subsection{System Architecture}
\label{ss:architecture}

\begin{figure}[tbp]
    \includegraphics[width=\linewidth]{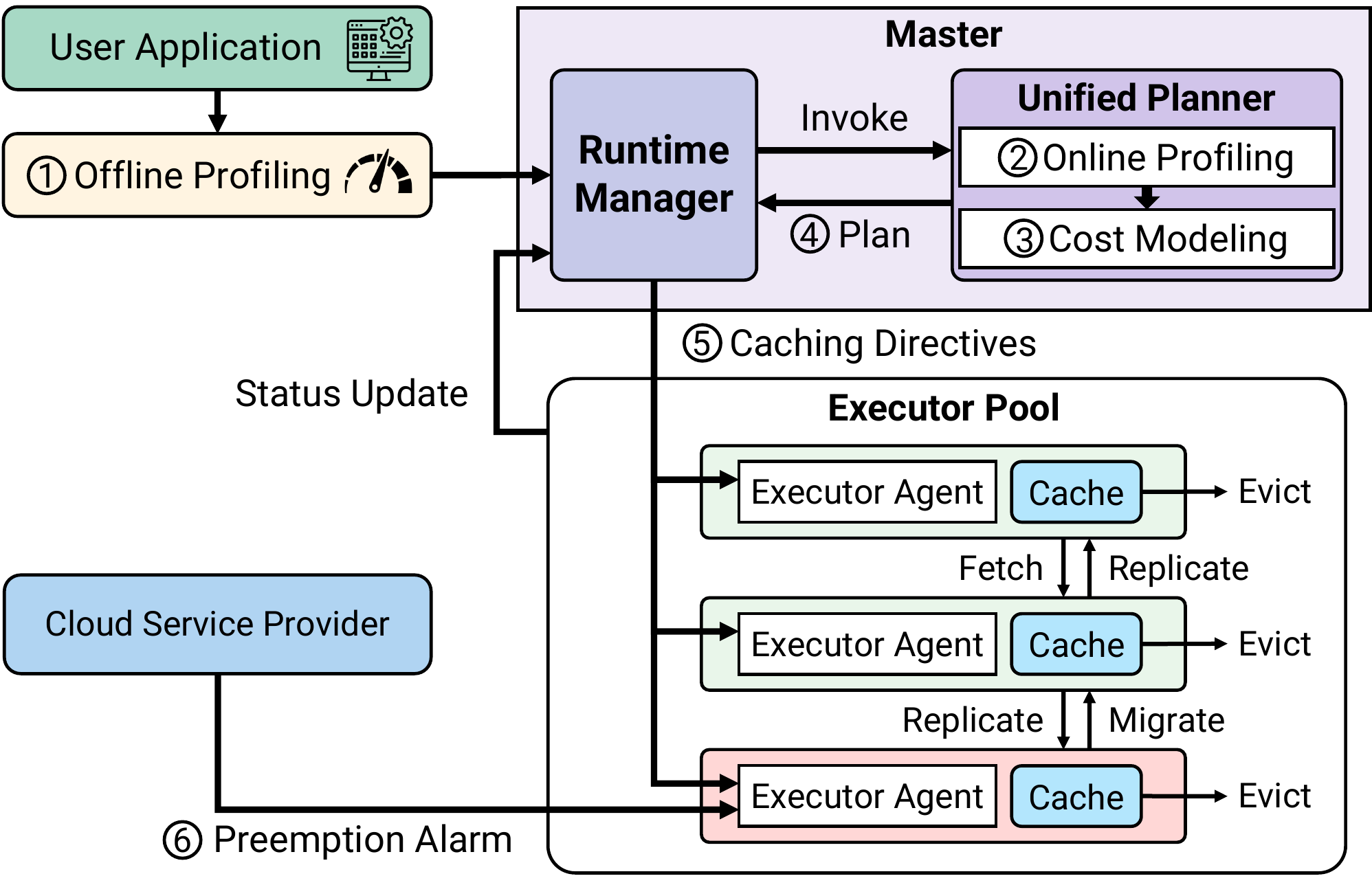}
    \caption{\label{fig:design-overview} Design Overview of \sys.}
\end{figure}

\autoref{fig:design-overview} illustrates the end-to-end operation of \ours
as a feedback loop between a centralized manager and distributed executors.
Before execution begins, \ours performs offline
profiling~\encircle{\textsf{1}} through a dry-run on a minimal input to
obtain the data dependency graph. During execution, executors report
the sizes and computation times of newly generated partitions along with
current cache states to the manager at each stage
boundary~\encircle{\textsf{2}},
keeping the profiled data current (\autoref{ss:profiling}). At each iteration
boundary, the cost model evaluates per-partition recovery costs across all
executors~\encircle{\textsf{3}} (\autoref{ss:cost-modeling}), and the MILP
solver consumes these costs to produce a cluster-wide caching
plan~\encircle{\textsf{4}} (\autoref{ss:milp}). The Runtime Manager diffs the
new plan against the current placement and distributes per-executor
directives~\encircle{\textsf{5}}. Executors then apply cache, evict, migrate,
and replicate actions, with execution mechanics described in
\autoref{s:implementation}. When a preemption alarm arrives or a new executor
joins~\encircle{\textsf{6}}, the manager re-invokes the solver under the
updated cluster composition
(\autoref{s:runtime-manager}).

\section{Unified Planner}
\label{s:cache-planning}

As established in \autoref{sec:overview}, \sys solves a
cluster-wide cache placement problem using a single MILP.
This section details the three components of \sys's
planning pipeline: input profiling,
cost modeling, and solver
design.

\subsection{Profiling}
\label{ss:profiling}

The solver can only produce an optimal plan if it
operates on accurate cost estimates, but partition
sizes, computation times, and cluster composition all
shift during execution.
\sys combines offline profiling to initialize these
estimates with online profiling to keep them up to date.

\PP{Offline Profiling}
Before execution begins, \sys performs a dry-run on a
minimal input to initialize all three state structures, similar to Blaze.
The dry-run measures initial partition sizes and
computation times to populate \textit{Cache State}, and
extracts the RDD lineage graph by capturing the code path
of the iterative computation to populate \textit{Cost Graph}.
Separately, \sys runs a bandwidth benchmark to obtain
the initial pairwise network bandwidth
$\texttt{BW}_n(e, e')$, per-executor disk bandwidth
$\texttt{BW}_d(e)$, and memory capacity $C_e$, which
together initialize \textit{Cluster State}.

\PP{Online Profiling}
\sys continuously updates all three
state structures during execution.
\textit{Cache State} and \textit{Cost Graph} are updated
continuously during execution while \textit{Cluster State} is updated on composition changes:
node departures and arrivals trigger an immediate bandwidth
benchmark so the solver incorporates the new cluster composition into
the next plan.

\subsection{Cost Modeling}
\label{ss:cost-modeling}

The cost model evaluates a caching plan by assigning each
partition a recovery cost that captures the expense
of making it available at the executor that needs it.
The MILP solver searches over caching plans to find the one
that minimizes total recovery cost across the cluster.

\PP{Caching Plan}
A caching plan specifies, for each partition $p$, one of
three storage states: cached in memory on one or more
executors, spilled to disk at its origin executor
$e_p$, or discarded entirely.
\sys uses two families of binary variables to encode this decision:
$x_{p,e} = 1$ indicates that executor $e$ holds $p$ in
memory, and $d_p = 1$ indicates that $p$ is spilled to
disk at $e_p$.
A partition is \emph{available} ($\alpha_p = 1$) whenever at least one
copy exists:
\begin{equation*}
  \alpha_p = \min\!\bigl(1,\;
    \textstyle\sum\limits_{e} x_{p,e} + d_p\bigr)
  \label{eq:avail}
\end{equation*}

\PP{Access Cost}
When an available partition $p$ of size $s_p$ is needed at executor
$e_t(p)$ 
(i.e., the target executor which does not already have $p$ in its cache)
, the access cost depends on where $p$ resides:
\begin{myenumerate}
  \item If $p$ is in memory at $e_t(p)$, the access
    cost is zero.
  \item If $p$ is in memory at a remote executor,
  the cost is a network transfer
    $s_p / \texttt{BW}_n(e,\, e_t(p))$.
    If multiple copies exist, \sys selects the
    copy with the highest bandwidth to $e_t(p)$.
  \item If $p$ is on disk at $e_t(p)$, the cost is a local disk read
    $s_p / \texttt{BW}_d(e_p)$.
  \item If $p$ is on disk at a remote origin
    $e_p \neq e_t(p)$, the cost compounds a disk read
    at $e_p$ with a network transfer to $e_t(p)$.
\end{myenumerate}
Using the convention $\texttt{BW}_n(e, e) = \infty$,
cases 1--2 and cases 3--4 each collapse into a single
expression.
$A(e_t(p))$ denotes the minimum access cost across all
locations where $p$ is available:
\begin{equation*}
  A(e_t(p)) =
  \begin{cases}
    \displaystyle
    \min_{e:\,x_{p,e}=1}
      (\frac{s_p}{\texttt{BW}_n(e,\, e_t(p))})
      & \text{if } \exists\, e:\, x_{p,e} = 1 \\[10pt]
    \displaystyle
    \frac{s_p}{\texttt{BW}_d(e_p)}
      + \frac{s_p}{\texttt{BW}_n(e_p,\, e_t(p))}
      & \text{if } d_p = 1
  \end{cases}
  \label{eq:access}
\end{equation*}
\PP{Recovery Cost}
When $p$ is \emph{unavailable} ($\alpha_p = 0$), it must be
recomputed at $e_t(p)$ from its parent partitions.
Each parent $q \in \texttt{parents}(p)$ is resolved
through the same cost analysis: if $q$ is \emph{available},
\sys pays the access cost $A_q(e_t(p))$; if $q$ is
unavailable, \sys recursively recomputes $q$ at $e_t(p)$.
Recovering multiple parents can proceed in parallel,
and prior work such as Blaze defines recovery cost as
the maximum parent cost to reflect wall-clock time on
a single executor.
However, in a cluster-wide view, each parent recovery
consumes executor resources regardless of parallelism:
parallel recovery at one executor comes at the expense
of work other executors could perform.
\sys therefore defines the recovery cost $R(e_t(p))$ as the sum over
all parents, capturing the total cost to the cluster:
\begin{equation}
  R(e_t(p)) = c_p +\textstyle\sum\limits_{q \in \texttt{par}(p)}
    \Bigl(\alpha_q  A(e_t(q))
    + (1 - \alpha_q)  R(e_t(q))\Bigr)
  \label{eq:recovery}
\end{equation}
where $c_p$ is the time to compute $p$ and $par(p)$ refers to the parents of partition $p$.
The recursion terminates at available partitions or
input data, where $R(e_t(p)) = 0$.

Note that this cost model fixes $e_t(p)$ (i.e., the executor scheduled to consume partition $p$) to the executor assignment
observed in the previous iteration and uses profiled computation times
without modeling the scheduler's reaction to caching decisions. Because \sys
places partitions across executors, its placement decisions influence where
locality-aware schedulers direct future tasks, coupling caching with
scheduling. Although this fixed assignment drifts as the scheduler
adapts---\autoref{fig:comp-skew} shows that load distributions shift between
consecutive iterations---recovery cost still captures the structural factors
that determine data importance, such as lineage depth, partition size, and
ancestor availability, which are intrinsic to the partition's position in
the DAG. Recovery cost under a fixed assignment is therefore a sound proxy
for prioritizing caching and placement decisions, even though it does not
predict the exact recovery time under any particular scheduler.

\subsection{Solver Design}
\label{ss:milp}
The solver's goal is to minimize the aggregate recovery cost of partitions globally, i.e., across all executors. As discussed in \autoref{ss:data-skew}, high recovery cost amplifies the memory and compute overhead on each executor when an eviction occurs. Minimizing the recovery cost reduces the resource overhead of handling the eviction of structurally important partitions, and addresses a critical bottleneck when locality-aware schedulers direct tasks to executors with most data. Scheduler decisions are not considered by the solver so that~\ours{} is broadly applicable to the class of locality-aware data analytic schedulers, without requiring specific scheduling policies to be modeled as they can vary across implementations. Doing so introduces the challenge of \emph{scheduler drift}, i.e., solver's cost estimates can drift from actual runtime costs when the scheduler assigns tasks differently. Such drift does not impact~\ours{}'s optimization decisions in a homogeneous resource setting.
We design \ours as a scheduler-agnostic caching layer,
so it can be integrated with existing locality-aware schedulers without
changing their task-assignment policies or modeling scheduler-specific
decisions.
The solver therefore determines the caching plan that minimizes
aggregate recovery cost alone, subject to memory and
placement constraints. The caching plan still prioritizes
structurally important partitions whose high recovery cost persists
regardless of scheduling outcome.
Let $\mathcal{J}$ denote the set of target partitions for
the next iteration---the partitions on which the computation
is first invoked, triggering the upstream computation
of parent RDD partitions.
The solver minimizes:
\begin{equation*}
  \underset{x_{p,e},d_p}{\text{minimize}} \quad
    \textstyle\sum\limits_{p \in \mathcal{J}} R_p(e_t(p))
  \label{eq:objective}
\end{equation*}

\PP{Constraints}
The objective is shared across all runtime scenarios, and \sys handles placement, preemption, and replication by activating different constraints for each case.
Across all scenarios, each executor's total cached data is bounded by its memory capacity $C_e$:
\begin{equation*}
  \textstyle\sum\limits_{p} s_p \cdot x_{p,e} \leq C_e,
  \quad \forall\; e
  \label{eq:capacity}
\end{equation*}
In the general case, each partition is either cached in memory on one executor, spilled to disk, or discarded:
\begin{equation}
  \textstyle\sum\limits_{e} x_{p,e} + d_p \leq 1,
  \quad \forall\; p
  \label{eq:exclusive}
\end{equation}
When a preemption alarm signals an executor departure, \sys excludes the departing executor $e_d$ from placement:
\begin{equation}
  \textstyle\sum\limits_{e \neq e_d} x_{p,e} \leq 1,
  \quad \forall\; p \in \mathcal{P}_{e_d}
  \label{eq:alarm}
\end{equation}
where $\mathcal{P}_{e_d}$ denotes the set of partitions cached on the departing executor.
To guard against unalarmed executor departures that
cause complete data loss without the overhead of full
checkpointing, \sys optionally replicates each shuffle
partition to $\rho$ executors, where $\rho$ is
user-configurable.
Shuffle partitions (denoted by $\mathcal{P}_s$) are singled out because their
recovery requires re-executing the preceding all-to-all
redistribution stage, making them the most expensive
partition type to recompute, while accounting for less than 1\% of the total cached data.
The origin executor always hosts one copy to preserve
data locality.
When enabled, the constraint for shuffle partitions
$\mathcal{P}_s$ overrides \autoref{eq:exclusive}:
\begin{equation}
  \textstyle\sum\limits_{e} x_{p,e} = \rho,
  \quad x_{p,e_p} = 1,
  \quad \forall\; p \in \mathcal{P}_s
  \label{eq:shuffle-rep}
\end{equation}
Although replicas add to memory pressure, this aligns with the cost-awareness principle: the solver makes room by evicting the lowest-cost partitions to retain the most expensive ones. For executor arrivals, \sys only updates \textit{Cluster State} to include the new executor.

\PP{Tractability}
Beyond the enlarged per-partition search space,
the cluster-wide formulation amplifies two structural
properties of the lineage graph that per-executor
solvers can largely ignore.
With the target executor of every
partition fixed, each of the $V$ partitions creates $O(E)$ variables and each
of the $L$ lineage edges a constant number of constraint terms, so the model
has size $O(VE+L)$. \sys applies one pruning technique for each to bound $V$
and $L$:
\begin{myenumerate}
  \item \textit{Cached-Block Boundary Pruning.}
    The summation over all parents in
    \autoref{eq:recovery} means that every ancestor's
    placement decision propagates to every descendant's
    recovery cost, not just descendants on the same
    executor or on the critical recomputation path.
    The full lineage, accumulated across all past
    iterations, therefore remains relevant to the solver: after $i$ iterations
    that each produce $k$ RDDs of $P$ partitions, $V = O(ikP)$. A cached
    partition serves as a recomputation barrier: no ancestor beyond it needs
    to be recomputed, so the solver removes those ancestors from the decision
    space and folds their cost into the cached partition. Only the lineage
    between $\mathcal{J}$ and the nearest cached partitions remains, so $V$
    scales with the depth of uncached lineage---typically an iteration or
    two---rather than with $i$.

\item \textit{Shuffle Approximation.}
    Shuffle outputs depend on every input partition
    through all-to-all redistribution, which makes them
    both the most expensive partitions to recover and
    the most expensive to model:
    the all-to-all dependency
    introduces $\Theta(P^2)$ edges into the lineage graph, a term that
    dominates $L$ as $P$ grows with the number of executors. In the cost graph
    fed to the solver, \sys removes the lineage between shuffle inputs and
    outputs: the cumulative cost of the full shuffle is assigned to each
    shuffle output, eliminating the $P^2$ input--output edges while retaining
    $P$ output-cost values, and the shuffle inputs are added to $\mathcal{J}$
    to approximate the severed lineage.
    This approximation keeps shuffle partitions
    prioritized by their high recovery cost while
    making the problem tractable.
    
\end{myenumerate}
Together, cached-block pruning bounds the lineage carried across
iterations, while shuffle approximation cuts the lineage at shuffle
boundaries, leaving only narrow dependencies so that $L = O(V)$. We evaluate
the resulting solver scaling in \autoref{fig:solver-scalability}.

\section{Runtime Manager}
\label{s:runtime-manager}

The Unified Planner determines the cache placement plan, while the Runtime
Manager determines how and when that plan is enforced. When the Planner
produces a new caching plan, the Runtime Manager diffs it against the most
recent plan and sends each executor the partition-level actions it must
apply: evict, migrate, or replicate. Only partitions whose state changes
appear in these directives. In response, executors report the sizes,
computation times, and residency states of newly created partitions to the
Runtime Manager at every stage completion, in order to keep \emph{Cache
State} and \emph{Cost Graph} up to date. This stage-level granularity is a
deliberate choice: per-task reporting would incur a round-trip for every
partition, while per-iteration reporting would prevent the Planner from
reacting promptly to mid-iteration composition changes such as preemptions.
\autoref{s:implementation} describes two execution mechanics of these
directives: lazy memory release and partition transfer optimization. This
section focuses on how the Runtime Manager adapts the caching plan to
executor arrivals, preemption alarms, and unalarmed departures.

\PP{Executor Arrival} By default, a newly joined executor provides no
immediate caching benefit because its memory is empty and cache population is
lazily coupled with tasks scheduled on that node.
\sys instead proactively profiles the new executor's network and disk
throughput, updates \emph{Cluster State}, and re-invokes the solver with the
expanded executor set.
The updated plan can redistribute partitions from memory-constrained
executors to the new node, exploiting the additional capacity immediately
rather than waiting for
task placement to populate the cache.

\PP{Preemption Alarm} The grace period between a preemption alarm and
instance reclamation varies across cloud providers, and not all cached
partitions can be migrated in time. Upon receiving an alarm, \sys re-invokes
the solver with~\autoref{eq:alarm} to account for
the departing executor. Once the Runtime Manager identifies the partitions to
migrate, \sys evacuates them in descending order of solver-estimated recovery
cost. Unlike prior cost-agnostic schemes such as the Spark Node
Decommissioner, which migrate partitions in arbitrary order, this ordering
ensures that the most expensive data is transferred first regardless of when
the instance is
reclaimed. Partitions computed during the grace
period also enter the evacuation queue as they become available. Any shuffle
partition that cannot be migrated in time remains available through proactive
replication, if enabled.

\PP{Unalarmed Departure} When an executor departs without an alarm, no
evacuation is possible and all locally cached data is lost. If proactive
replication is enabled, shuffle partitions survive through replicas already
residing on other executors.
\sys exposes this replication as a user option, since unalarmed departures
are rare. Otherwise, all lost partitions are recovered through lineage
recomputation when a subsequent task requires them.
As with an alarmed departure, the Runtime Manager re-invokes the solver at
the next iteration boundary with the reduced executor set to produce an
updated plan.

\section{\sys Implementation}
\label{s:implementation}

\sys is implemented on top of Apache Spark 3.3.2 by modifying approximately
8k lines in the storage and scheduling subsystems. On the master,
\cc{\name~RpcEndpoint} coordinates cluster-wide caching decisions and
maintains \cc{RDDJobDag}, which tracks RDD dependencies, job-to-stage
mappings, and partition-level metrics collected from all executors. Executors
report partition sizes and computation costs to the master via asynchronous
RPC messages, and \cc{RDDJobDag} aggregates these reports into per-partition
cost estimates for the solver. On each executor, \cc{\name~Manager}
intercepts caching and eviction decisions within \cc{BlockManager} through a
pluggable \cc{CostPolicy} interface, which maps solver outputs to
per-partition actions: cache, disk, transfer, or evict. When a solver result
is available, \cc{\name~Manager} consults the placement map to determine
whether a partition belongs on the local executor; otherwise, it falls back
to reuse-based autocaching. Partition migration and replication use Spark's
\cc{BlockTransferService}, with the solver's per-partition replication factor
determining the number of replicas. On a preemption signal, the master marks
the alarmed executor, removes it from the active set, triggers the solver
under the updated configuration, and broadcasts the new plan to surviving
executors via RPC.

The \sys solver runs as a standalone Python process using Gurobi
11.0.3~\cite{gurobi}, communicating with the JVM master through named pipes
in JSON format. The master serializes the compressed DAG, per-executor memory
capacities, and measured bandwidth metadata to the pipe, and the solver
formulates the MILP and writes back the block-to-executor assignment.

\PP{Lazy Memory Release} When generating a caching plan, the solver considers
both resident partitions and new partitions to be produced in the next
iteration. Because new partitions materialize progressively, eagerly
reclaiming memory for partitions already transferred to remote executors
would leave local memory underutilized. \ours therefore marks such partitions
as removable and defers reclamation until the executor's memory is
exhausted.
Once memory is full, the executor
frees removable partitions in ascending order of recovery cost.

\PP{Partition Transfer Optimization} \ours performs partition transfers
through three mechanisms: asynchronous transfer, batching, and pipelining.
Each executor buffers partitions targeting the same remote executor and
flushes them in batches once the buffer is full to reduce per-transfer
overhead. It further pipelines transfers across all destinations, as the CPU
overhead of asynchronous transfers is low.
Since the shuffle stage
saturates network bandwidth with all-to-all communications, remote transfers
pause until it completes. Replication follows the same mechanism, except that
the sender retains its copy permanently.

\section{Evaluation}
\label{sec:evaluation}
\label{sec:caching-eval}

\subsection{Experimental Setup}

\PP{Environment} Unless specified otherwise, we deploy \ours{} on 11
\texttt{r5n.2xlarge} AWS EC2 instances (8~vCPUs, 64~GB memory, 512~GB gp3
SSD, up to 25~Gbps network), with one master and ten worker nodes hosting
20~executors of 25~GB memory each. We use Spark's default memory management
settings~\cite{spark-config}, under which execution and caching dynamically
share a single memory region of 318~GB in aggregate.

For the full-disaggregation comparison in
\autoref{sec:design-validation}, we keep executor CPU and execution memory allocations
identical across both configurations. For full-disaggregation setting, we use 
Amazon S3~\cite{awss3} to hold cached data in dedicated cloud object
storage. This setting uses five \texttt{m6in.\allowbreak{}16xlarge} workers 
with 256~GiB of memory each, while \ours uses five \texttt{r6in.\allowbreak{}16xlarge} workers with
256~GiB for execution memory, and 256~GiB for storage memory (i.e., for caching) each. 
Both instance types provide 64~vCPUs and 100~Gbps network. Each worker runs four 8-core executors.

\PP{Benchmarks} We evaluate four benchmarks spanning graph analytics and
machine learning, selected from those evaluated in~\cite{blaze}.
\textbf{PageRank~(PR)} iteratively propagates influence scores across a
graph, producing a deep lineage and the largest working set among our
benchmarks: each iteration references about 400~GB of intermediate
partitions, 1.3$\times$ the aggregate cache. \textbf{Connected
Components~(CC)} identifies disjoint subgraphs of mutually reachable vertices
on the same input graph. It shares PR's iterative structure but converges in
fewer iterations, producing shorter lineage chains and a working set of
276--383~GB per iteration. Both operate on a 54~GB synthetic graph that
contains 25~million vertices (SparkBench~\cite{sparkbench}). \textbf{Logistic
Regression~(LR)} optimizes binary classification coefficients on a 118~GB
click-log dataset (HiBench~\cite{hibench}). Its computational graph is
compact, and its working set stays close to the dataset size and fits within
the cache. \textbf{SVD++} performs collaborative filtering on synthetic
ratings from 15M users across 1.5M items (25~GB). Its working set
(264--384~GB) is smaller than PR's, but each partition carries high recovery
cost from heavier serialization.

\begin{figure*}[t]
  \centering
  \edef\vmfighA{\the\dimexpr0.14787\textwidth\relax}%
  \edef\vmfighB{\the\dimexpr0.16243\textwidth\relax}%
  \begin{minipage}[t]{0.4962\linewidth}
    \vspace{0pt}\centering
    \includegraphics[height=\vmfighA,trim=1.5bp 2.5bp 1.5bp 2.57bp,clip]{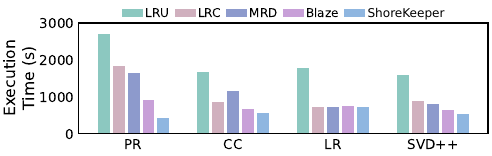}
  \end{minipage}
  \hfill
  \begin{minipage}[t]{0.4584\linewidth}
    \vspace{0pt}\centering
    \includegraphics[height=\vmfighB,trim=6.5bp 6.7bp 6.5bp 14.64bp,clip]{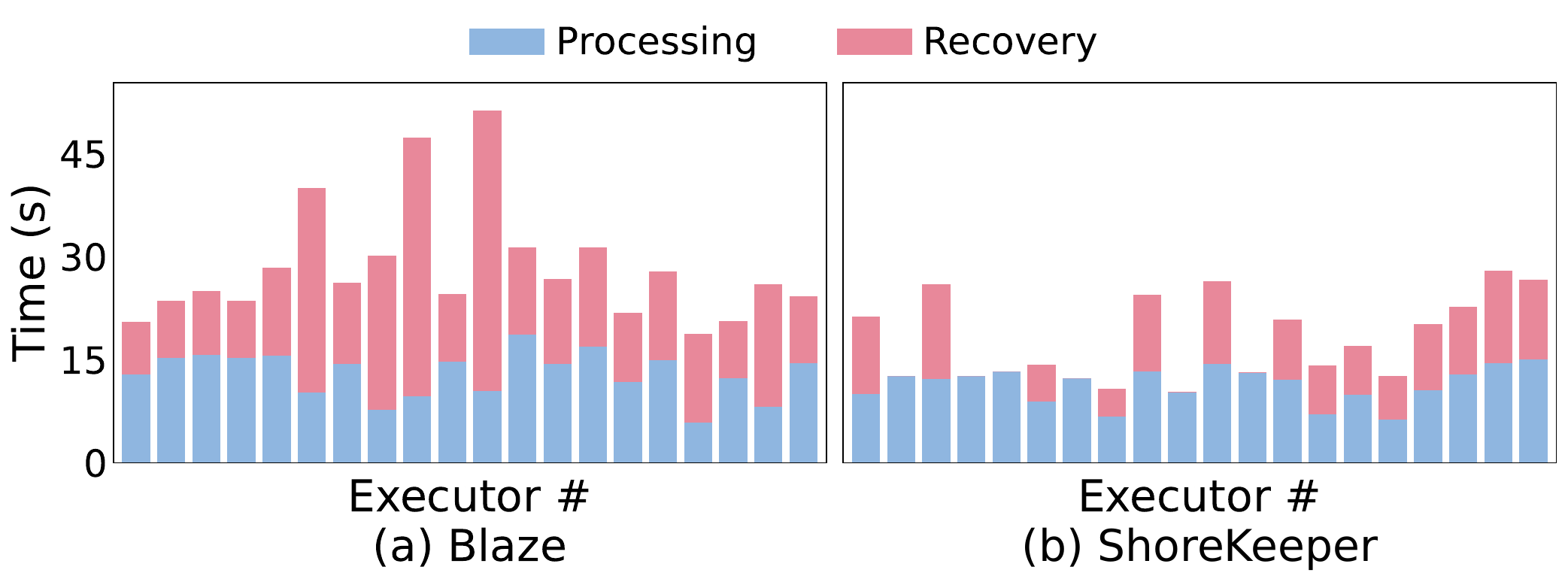}
  \end{minipage}
  \par
  \begin{minipage}[t]{0.4962\linewidth}
    \caption{Performance comparison on VM instances.}
    \label{fig:e2e_comparison}
  \end{minipage}
  \hfill
  \begin{minipage}[t]{0.4584\linewidth}
    \caption{Per-executor processing and recovery costs
    under Blaze and \ours at PR iteration 5 on VM instances.}
    \label{fig:load_balancing}
  \end{minipage}
\end{figure*}

\begin{figure*}[t]
    \centering
    \includegraphics[width=\linewidth]{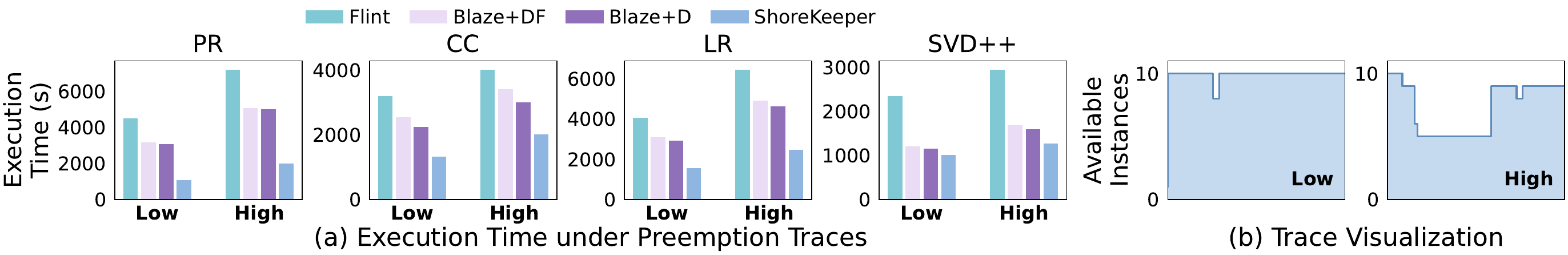}
    \caption{(a) Execution time comparison under
    preemption traces (\textit{Low},
    \textit{High}).
    (b) Visualization of preemption trace segments.}
  \phantomsubcaption\label{fig:preemption-main}
  \phantomsubcaption\label{fig:trace_visualization}
\end{figure*}

In PR, CC, and SVD++, the cache is thus full throughout the iterations and
47--75\% of cached partitions are evicted before a job completes.

\PP{Baselines} We compare \ours against five baselines categorized into three
groups: heuristic-based caching (LRU, LRC~\cite{lrc}, and MRD~\cite{mrd}),
cost-aware caching (Blaze~\cite{blaze}) and in-storage checkpointing
(Flint~\cite{flint}).

\begin{myitemize}
  \item \textbf{LRU} is the default caching strategy in vanilla Spark that
  relies on manual annotations for caching decisions,
  following the standard Apache Spark
  GraphX~\cite{graphx} and Spark MLlib~\cite{mllib} implementations for the corresponding benchmarks.

  \item \textbf{LRC and MRD} are representative cost-agnostic policies.
  LRC (Least Reference Count) evicts partitions with the fewest references in the compute graph.
  MRD (Most Reference Distance) evicts partitions farthest from their next access and
  prefetches upcoming data.

  \item \textbf{Flint} is a checkpointing-based system that periodically
  persists frontier partitions, the latest outputs at the leading edge of
  the lineage, to network-attached block storage.

  \item \textbf{Blaze} is a
  recovery cost-aware caching system that
  solves a per-executor ILP for caching decisions (\autoref{eq:blaze-objective}).
\end{myitemize}

All baselines are implemented in the same Spark engine for a fair comparison.
Except for Flint, none of the baselines support preemption handling. We thus
build two additional preemption-aware baselines: (1) \textbf{Blaze+D}, which
integrates the Spark Node Decommissioner~\cite{spark-decommission} into Blaze to
migrate randomly selected cached blocks to surviving peers on a preemption
alarm, and (2) \textbf{Blaze+DF}, which further adds Flint's periodic
checkpointing atop Blaze+D for durable replication.

\subsection{Deployments on VM Instances}

\autoref{fig:e2e_comparison} compares the job completion time of \ours and
all baselines across four workloads on stable on-demand instances. Overall,
\ours achieves 2.05--6.11$\times$, 1.20--3.03$\times$,
1.01--2.40$\times$, and 1.19--2.99$\times$ speedup over the 
baselines in PR, CC, LR, and SVD++,
respectively. Among all workloads, PR presents the highest cross-executor
imbalance and the deepest lineage, making it the most challenging caching
scenario. \ours achieves its largest performance gain here, demonstrating the
benefit of cluster-level placement under heavy memory pressure. LRU in
vanilla Spark shows the longest execution time across workloads.

On PR, Blaze outperforms MRD by 1.8$\times$ through
cost-awareness, yet per-executor optimization still reaches a ceiling
even when the policy is cost-aware.
\ours overcomes this limit by coordinating placement across executors.
On CC, \ours achieves a moderate speedup over Blaze. CC's shorter lineage
reduces the recomputation penalty per evicted partition and leaves less room
for cluster-level placement.
SVD++ shows a gap similar to CC, but for a different reason: its high
per-partition recovery cost amplifies the penalty of suboptimal eviction.
This gives \ours
an opportunity to redirect expensive partitions to executors with spare
memory. For LR, the gap between all methods narrows considerably, since LR's
compact computational graph and small working set fit within per-executor
memory. Evictions thus occur rarely regardless of the caching strategy.

\PP{Recovery Cost Balancing} \ours reduces end-to-end time by balancing
recovery cost across executors. \autoref{fig:load_balancing} thus compares
per-executor processing and recovery cost between Blaze and \ours at the 5th
iteration of PR. Under Blaze, recovery cost concentrates on a few executors
while others remain lightly loaded. \ours flattens this distribution by
placing high-cost partitions on executors with spare capacity. We also
observe a slight reduction in per-executor processing time under \ours. This
is likely a coupled effect: when busy executors hold a disproportionate share
of partitions, Spark's locality-aware scheduler delays task launches waiting
for a local slot before falling back to a remote executor.
By proactively redistributing partitions to lighter-loaded executors,
\ours makes local slots more readily available and reduces scheduling delays.

\subsection{Deployments on Preemptible Instances}
\label{sec:preemption-eval}

To evaluate \ours under realistic resource variability, we collect traces
from 10 AWS EC2 r5-family Spot Instances in \texttt{us-east-1} over six days
and extract two windows with distinct preemption frequencies
(\autoref{fig:trace_visualization}). The first window (\emph{Low}) captures a
relatively stable period with sporadic single-instance preemptions. The
second window (\emph{High}) reflects intensive interruptions that shrink the
cluster to 5~nodes and includes one unalarmed departure. Each
benchmark spans a different range within these windows due to varying
execution times.

\begin{figure}[!tb]
  \centering
  \includegraphics[height=0.16309\textwidth,trim=4bp 12.44bp 1bp 5.5bp,clip]{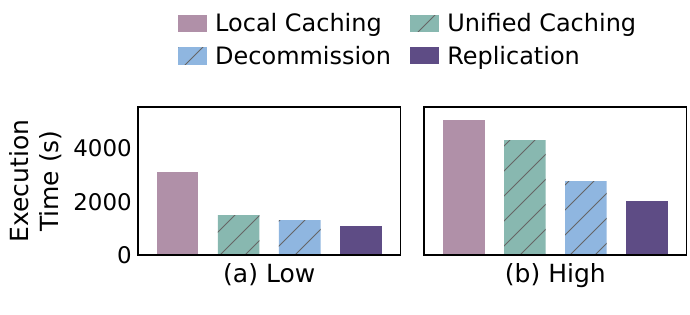}
  \caption{Component breakdown for PageRank under
  (a) Low and (b) High, adding components incrementally to local caching.}
  \label{fig:perf_breakdown}
\end{figure}

\autoref{fig:preemption-main} compares the job completion time of \ours
against Flint, Blaze+DF, and Blaze+D under both windows. \ours matches or
outperforms every baseline across all workloads.
Overall, \ours achieves 1.14--2.87$\times$ speedup over Blaze+D
depending on workload and preemption frequency.
The largest gap appears on PR, reaching 2.87$\times$ over Blaze+D under Low
and 2.5$\times$ under High.
Such a gap arises mainly because Blaze+D migrates cached partitions during
the grace period without prioritizing them by recovery cost. \ours instead
invokes its lightweight solver to evacuate partitions based on recovery cost,
and proactively replicates shuffle partitions during normal execution
(replication factor set to two in our experiments;
\autoref{fig:perf_breakdown} breaks down each effect). Notably, \ours retains
its advantage under High, where multiple instances are preempted
simultaneously and the grace period allows little time to react.

For CC and SVD++, the performance gap between \ours and Blaze+D is narrower
than in PR. Lower memory pressure and less imbalance in recovery cost
make each preemption cheaper to recover from, reducing the advantage that
\sys provides. LR is a notable case: its per-executor memory pressure is
negligible on stable instances, yet it still shows a gap under preemption.
Preemption can thus cause the loss of critical cached data regardless of
memory pressure.

Flint incurs the longest completion time across all workloads and both
windows, as its periodic checkpointing to HDFS adds I/O overhead on every
iteration regardless of whether preemption occurs. Blaze+DF likewise provides
no consistent improvement over Blaze+D, as the I/O cost of checkpointing
outweighs the recomputation it avoids.

\begin{figure}[t]
  \centering
  \includegraphics[width=\columnwidth,trim=6bp 2.36bp 2.5bp 1.5bp,clip]{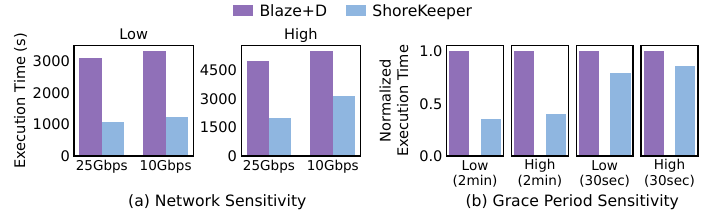}
  \caption{
      (a) Execution time on PageRank at 10~Gbps and 25~Gbps
      (b) Normalized execution time on PageRank under Low and High with different grace periods.
  }
  \phantomsubcaption\label{fig:network-preemption}
  \phantomsubcaption\label{fig:preemption-alarm-sensitivity}
\end{figure}

\subsection{Sensitivity Studies}

We evaluate the sensitivity of \ours's performance to variations in network
bandwidth, grace period duration, and deployment scale.

\PP{Network Bandwidth} Slower networks penalize both shuffle execution
and remote recomputation, and \ours adds migration traffic competing for the
same bandwidth. We therefore run \ours and Blaze+D on \texttt{r5.2xlarge}
instances (up to 10~Gbps, compared to 25~Gbps on r5n) under both preemption
windows (\autoref{fig:network-preemption}). Both systems slow down relative
to r5n, but \ours maintains a consistent gap over Blaze+D: the recomputation
saved by cluster-level placement outweighs the additional migration cost.

\PP{Grace Period Duration} \autoref{fig:preemption-alarm-sensitivity}
compares \ours and Blaze+D under two grace periods, 2~mins and 30~secs, with
execution time normalized to Blaze+D. A shorter window penalizes \ours more
than Blaze+D, because \ours evacuates partitions in descending order of
recovery cost and thus benefits more from a longer window. In contrast,
Blaze+D is cost-agnostic and gains little from the extra time. Even so, \ours
retains its advantage at 30~secs, since the partitions it transfers are the
most expensive ones to recompute.

\begin{figure}[t]
  \centering
  \includegraphics[width=\linewidth]{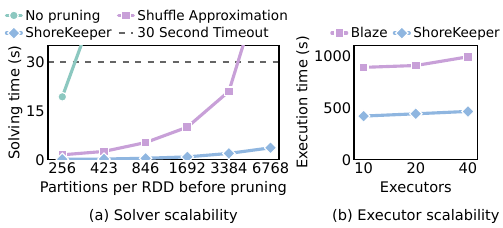}
  \caption{Scalability on PageRank. (a) Solver time as the
  partition count per RDD increases. (b) Execution time at 10, 20, and 40
  executors with input scaled by 0.5$\times$, 1$\times$, and
  2$\times$.}
  \label{fig:scalability}
  \begin{subcaptiongroup}
    \phantomcaption\label{fig:solver-scalability}
    \phantomcaption\label{fig:executor-scalability}
  \end{subcaptiongroup}
\end{figure}

\PP{Solver Scalability} Pruning must keep solving time tractable as workloads
grow. To assess its efficacy, we stress the solver under different pruning
settings by scaling PR's default 423 partitions per RDD, measured at
iteration 10, up to 16$\times$. \autoref{fig:solver-scalability} shows that
\ours stays below 4~secs at 6,768 partitions per RDD (16$\times$ the original
scale), while no pruning exceeds 30~secs at 423 partitions and shuffle
approximation exceeds 30~secs at 6,768. Our pruning techniques are therefore
necessary at scale, yet they incur little optimality loss: across all 
four workloads, the recovery-cost objective stays within 2\% of that without pruning.

\PP{Executor Scalability} We also evaluate executor scaling on PR by
extending the setting from \autoref{fig:e2e_comparison} to 10 (0.5$\times$
input), 20 (1.0$\times$ input), and 40 executors (2.0$\times$ input), with per-executor resources
fixed. As shown in \autoref{fig:executor-scalability}, \ours remains roughly
$2.1\times$ faster than Blaze, with only a 10.7\% increase in execution time
as the system scales.

\subsection{Design Validation}
\label{sec:design-validation}

We isolate individual design choices to validate their contribution to
\ours's performance.

\PP{Performance Breakdown} \autoref{fig:perf_breakdown} incrementally
enables \ours's components on PR under both preemption windows, with local
caching matching Blaze. Under Low, unified caching halves execution time and
accounts for around 80\% of the total speedup. Under High, all three
components contribute substantially, as cost-aware decommission and proactive
replication are necessary to cope with frequent node preemptions.

\PP{Runtime Overhead} \label{sec:overhead-analysis} \ours introduces
three sources of runtime CPU overhead, all insignificant across the four
workloads: profiling at 1.4--2.4\%, solver below 2.7\%, and migration threads
below 0.05\%.

\PP{In-Memory Caching vs. Full Disaggregation} AnyBlob~\cite{anyblob}
motivates direct object-store reads over fast networks as an alternative to
in-memory caching. We test this in Spark by writing every partition selected
for local caching to S3 and reading it back on demand, using workers with
half the physical memory at 20\% lower instance cost. Over ten PR iterations
on a 100M-vertex graph, such full disaggregation takes 2.02$\times$ longer
and incurs 66.9\% higher estimated cost
(\autoref{tab:disaggregated-state-cost}). The comparison excludes EBS costs,
which both systems incur equally, and S3 request fees, which apply only to
full disaggregation.

\section{Discussions}
\label{sec:discussion}

\PP{Profiling Overhead} Offline profiling provides initial estimates that \ours refines online during execution. 
Its startup cost amortizes across long iterative jobs but can matter
for short jobs. Sampling or reusing profiles from similar workloads could reduce this cost.

\PP{Deployment Scenarios} When all instances are simultaneously reclaimed in
a purely preemptible cluster, \ours falls back to full recomputation from the
original input. A simple remedy is to provision a small number of on-demand
nodes alongside preemptible ones, a configuration that \ours handles without
modification. For a cluster with on-demand instances, node departures can
occur through hardware failures or autoscaling, which resemble unalarmed and
alarmed departures, respectively. \ours covers both cases through the same
Runtime Manager mechanisms.

\PP{Generality}
\ours requires three properties from the backend: intermediate data
materialized as immutable partitions, a dependency graph over those
partitions, and the ability to recompute a lost partition from its ancestors.
Engines that satisfy these properties (e.g., Flink~\cite{flink}, Ray~\cite{ray}, and
Dask~\cite{dask}) could serve as a backend with modest engineering effort, as each
component of \ours operates on this abstraction.
Extending beyond iterative workloads is harder: iteration boundaries bound
the solver's scope to the next iteration, and without them it must reason
over the entire dataflow at once.

\begin{table}[t]
  \centering
  \captionsetup{position=bottom,skip=4pt}
  \small
  \renewcommand{\arraystretch}{0.92}
  \begin{tabular*}{\linewidth}{@{\extracolsep{\fill}}lrr@{}}
    \toprule
     & \textbf{\ours} & \textbf{Full S3} \\
    \midrule
    Workers (\$/h) & $5\times 5.5786$ & $5\times 4.4554$ \\
    Master (\$/h) & 0.697 & 0.697 \\
    \midrule
    Hourly total (\$/h) & 28.590 & 22.974 \\
    Execution time (s) & 1,900.17 & 3,847.83 \\
    \textbf{Total (\$)} & \textbf{15.09} & \textbf{25.18} \\
    \bottomrule
  \end{tabular*}
  \normalsize
  \caption{Execution time and cost for ten PageRank iterations.
  Full S3 total includes \$0.6241 of storage. Both exclude EBS and S3 request fees.}
  \label{tab:disaggregated-state-cost}
\end{table}

\section{Related Work}
\label{sec:relwork}

\PP{Distributed Memory Management}
Memory disaggregation enables applications to access
remote memory over the network. FaRM~\cite{farm1} and soNUMA~\cite{sonuma}
expose cluster memory as an explicit shared address space over RDMA for
low-latency access. Infiniswap~\cite{NSDI17_InfiniSwap} and
Fastswap~\cite{fastswap} harvest idle remote memory through RDMA-backed
paging. AIFM~\cite{aifm} and Semeru~\cite{semeru} integrate far memory
management into application runtimes to hide far-memory access latency.
Carbink~\cite{carbink} features fault tolerance through erasure coding and
ownership migration. These systems treat remote memory as an
extension of local memory, whereas \ours manages intermediate data
placement across dynamically changing cluster nodes to reduce recovery cost.

\PP{Adapting to Resource Changes} Several systems adapt computation or its
state to changing cloud resources. Pado~\cite{pado} schedules critical tasks
on reserved instances while offloading elastic work to preemptible ones, and
ExoSphere~\cite{exosphere} manages portfolios of heterogeneous transient
instances to optimize availability and cost. Accordion~\cite{accordion}
changes the parallelism of an analytical query during execution and rebuilds
partitioned state as resources change. BigQuery's distributed
shuffle~\cite{bigqueryshuffle} separates shuffle data from compute workers,
while
POLARIS~\cite{polaris} separates compute from state by externalizing the
assignment of data cells to nodes, enabling partial query restart and online
resize.
On-demand state separation~\cite{ondemandstate} implements task migration
without costly periodic checkpointing by extracting and transferring operator
state and execution progress only when migration is triggered.
These systems reconfigure execution state to survive resource changes, but
treat all state as equally worth preserving. \ours instead decides which
reusable data partitions to retain based on their lineage-aware recovery
costs.

\PP{Checkpoint-Based Recovery}
Besides Flint~\cite{flint}, several systems persist intermediate state to
stable storage to survive failures or instance revocations, including
TR-Spark~\cite{trspark}, SpotOn~\cite{spoton}, Tachyon~\cite{tachyon},
and cost-based approaches~\cite{daly, ftopt, salama-ft, phoebe}.
These systems trade storage and I/O cost for bounded recovery time,
whereas \sys retains intermediate data in memory, optimizing caching,
placement, and replication as cluster capacity changes.

\PP{External Storage and Analytical Caching} A separate line of work moves
analytical data and state off compute workers. Among analytical caches,
Crystal~\cite{crystal} places a cache next to compute and stores
predicate-defined regions of base data to reduce reads from remote cloud
storage, and AnyBlob~\cite{anyblob} retrieves base data from cloud object
stores through a download path integrated with the query engine. Among
disaggregated stores, Pocket~\cite{pocket} provides job-scoped ephemeral
storage, Locus~\cite{locus} spreads serverless shuffle data across object
storage and a fast storage tier, and Jiffy~\cite{jiffy} leases fine-grained
far memory to serverless tasks. Tachyon~\cite{tachyon}, later renamed
Alluxio, uses storage-level lineage for high-throughput writes and recovery
while exposing a shared file layer to compute frameworks such as
Spark~\cite{alluxiospark}. These systems change where data resides and how it
is accessed, while \ours decides which reusable partitions to retain as
executor capacity changes. The two operate at different layers of the caching
stack and can be combined.

\section{Conclusion}
\label{sec:conclusion}

Workload analysis reveals that managing recovery
costs across executors in the cloud for iterative workloads is essential for reducing execution
time. \ours{} addresses two intrinsic challenges:
recovery cost
imbalance across executors and resource preemption, making it applicable across diverse cloud deployment scenarios. Its design
follows a clean separation of planning and execution, where the planner uses MILP to holistically balance caching benefits against eviction
costs across all in-memory partitions in the cluster.

\bibliographystyle{ACM-Reference-Format}
\bibliography{refs}

\end{document}